\documentstyle[graphicx]{article}

\begin{document}

\author{E. J. Calegari$^a$, S.G. Magalh\~{a}es$^a$ and A.A. Gomes$^b$ \and {\it $^a$Departamento de
F\'{\i}sica-UFSM,97105-900 Santa Maria, RS, Brazil} \and %
{\it $^b$Centro Brasileiro de Pesquisas F\'{\i}sicas, Rua Xavier Sigaud 150,}\\
{\it 22290-180, Rio de Janeiro, RJ, Brazil}}
\title{Role of Hybridization in the Superconducting Properties of an
Extended $d-p$ Hubbard Model: a Detailed Numerical Study }
\maketitle

\begin{abstract}
The Roth's two-pole approximation has been used by the present
authors to study the effects of the  hybridization in the
superconducting properties of a strongly correlated electron
system. The model used is the extended Hubbard model which
includes the $d-p$ hybridization, the $p$-band and a narrow
$d$-band. The present work is an extension of the previous Ref.
\cite{ref3}. Nevertheless, some important correlation functions
necessary to estimate the Roth's band shift, are included together
with the temperature $T$ and the Coulomb interaction $U$ to describe
the superconductivity. The superconducting order parameter of a
cuprate system, is obtained following Beenen and Edwards
formalism. Here, we investigate in detail the change of the order
parameter associated to temperature, Coulomb interaction and
Roth's band shift effects on superconductivity. The phase diagram
with $T_c$ versus the total occupation numbers $n_T$, shows the
difference respect to the previous work.
\vspace{.4cm}
\\
PACS numbers: 71.27.+a, 71.10.Fd, 74.25.Dw


\end{abstract}

\vspace{.8cm}

The high temperature superconductors discovered by Bednorz and Muller
\cite{ref1} are believed to be explained in the framework of
the Hubbard model \cite{ref2} since electron correlations are strong.

In this work we have used the extended Hubbard model \cite{ref3}
with the Roth's method \cite{ref4}. In the previous work
\cite{ref3}, we improved Ref. \cite{ref5} to also include
superconducting properties following closely the approach
introduced by Beenen and Edwards \cite{ref6}. Nevertheless, in
Ref. \cite{ref3}, the Roth's band shift has been estimated
disregarding the temperature and the Coulomb interaction effects.
In the present work, we have included these effects in the Roth's
band shift. Consequently, due to the $U$ dependence, it is necessary
to calculate more correlation functions than in the reference
\cite{ref3}. Therefore, we hope by using this procedure to get a
more correct behavior of the superconductor order parameter (as a
function of $T$) and the critical temperature $T_c$. The Hamiltonian
used is:
\begin{eqnarray}
H&=&\sum_{i,\sigma }(\varepsilon _{d}-\mu)d_{i\sigma
}^{\dag}d_{i\sigma }+\sum_{i,j,\sigma }t_{ij}^{d}d_{i\sigma
}^{\dag}d_{j\sigma }+
U\sum_{i}n_{i\uparrow}^{d}n_{i\downarrow}^{d}\nonumber\\
 & &+\sum_{i,\sigma }(\varepsilon _{p}-\mu)p_{i\sigma }^{\dag}p_{i\sigma
}+\sum_{i,j,\sigma }t_{ij}^{p}p_{i\sigma }^{\dag}p_{j\sigma }\nonumber\\
 & &+\sum_{i,j,\sigma }t_{ij}^{pd}\left( d_{i\sigma
}^{\dag}p_{j\sigma +}p_{i\sigma }^{\dag}d_{j\sigma }\right)
\label{eq1}
\end{eqnarray}
where $\mu$ is the chemical potential.

In order to study superconductivity by Roth's method, it is
necessary to include "hole" operators in the set of "electron"
operators (which describes the normal state) and evaluate
anomalous correlation functions. As discussed in Ref. \cite{ref3},
considering the particular case with singlet pairing and the
$d$-wave symmetry, we get $\langle d_{i-\sigma }d_{i\sigma
}\rangle = 0 $ and $\sum_l\langle d_{i-\sigma }d_{l\sigma }\rangle
= 0 $, where the sum is over sites $l$ which are nearest neighbors
of $i$. Therefore, in the $d$-wave case \cite{ref6}, the
superconducting gap is determined by the gap function,
\begin{eqnarray}
\overline{\gamma}_{k}&=&-\overline{\gamma}_{k}\frac{2n_{1\sigma}^dt^{d}U^2}{L}
\sum_{k}\left[\cos{(k_xa)}-\cos{(k_ya)} \right]^2\frac{1}{2\pi i}{\displaystyle \oint} f(\omega)
G_{k\sigma }^{13}(\omega)d\omega
\label{eq2}
\end{eqnarray}
where $f(\omega)$ is the Fermi function. The propagators
$G_{k\sigma }^{1s}$, with $s=1,2$ or $3$ can be obtained as in
Ref. \cite{ref3}. We can readily obtain the correlation function
$n_{1\sigma}^d=\langle d_{1\sigma }^{\dag}d_{1\sigma }\rangle$
considering a nearest-neighbor model with $t_{0j}^{d}=t^{d}$ for
the $z$ neighbors. In this way it is necessary only one value of
$n_{j\sigma}^d$, which is $n_{1\sigma}^d$. The gap function given
by equation (\ref{eq2}), defined in Ref. \cite{ref6}, is solved
self-consistently. The band shift $W_{k\sigma}=W_{k\sigma}^d +
W_{\sigma}^{pd}$, shifts the poles of the propagators $G_{k\sigma
}^{1s}$. In this work we considered a $k$ independent
hybridization $(V_0^{pd})^2=\langle V_k^{dp}V_k^{pd}\rangle$ as
discussed in Ref. \cite{ref5}, where $\langle ...\rangle$ is the
average over the Brillouin zone and $V_k^{dp}(V_k^{pd})$ are the
Fourier transform of $t_{ij}^{dp}(t_{ij}^{pd})$. Consequently, the
$W_{\sigma}^{pd}$ term is $k$ independent, and also, its
temperature and Coulomb interaction dependence is quite small.
Thus, our focus here is $W_{k\sigma}^{d}$ which is given by:
\begin{eqnarray}
W_{k\sigma}^d&=&-\frac{1}{n_{\sigma}^d(1-n_{\sigma}^d)}\sum_{j\neq 0}t_{0j}^{d}\left\{
               (n_{j\sigma}^d-2m_{j\sigma})+e^{i{\bf{k}} \cdot {\bf{R}}_j}
            \left[ \frac{\alpha_{j\sigma}n_{j\sigma}^d+\beta_{j\sigma}m_{j\sigma}}
	    {1-\beta_{\sigma}\beta_{-\sigma}}\right.\right.\nonumber\\ 
	    & &\left.\left.+ \frac{\alpha_{j\sigma}n_{j-\sigma}^d+\beta_{j\sigma}m_{j-\sigma}}{1+\beta_{\sigma}} 
            +\frac{\alpha_{j\sigma}n_{j-\sigma}^d+\beta_{j\sigma}(n_{j-\sigma}^d-m_{j-\sigma})}
           {1-\beta_{\sigma}}\right] \right\}
\label{eq5}
\end{eqnarray}
where $\alpha$ and $\beta$ are defined as:
\begin{equation}
\alpha_{j\sigma}=\frac{n_{j\sigma}^d-m_{j\sigma}}{1-n_{-\sigma}^d}~~~
\mbox{and}~~~
\beta_{j\sigma}=\frac{m_{j\sigma}-n_{-\sigma}^dn_{j\sigma}^d}{n_{-\sigma}^d(1-n_{-\sigma}^d)},
\label{eq6}
\end{equation}
with $n_{-\sigma}^{d}\equiv n_{0-\sigma}^{d}$. The correlation
function $m_{j\sigma}=\langle d_{0\sigma }^{\dag}n_{j-\sigma}^d
d_{j\sigma }\rangle$ is obtained from the propagator $G_{k\sigma
}^{12}$ (see Ref. \cite{ref3}) as:
\begin{equation}
m_{j\sigma}=\frac{1}{2\pi i}{\displaystyle \oint} f(\omega)
\frac{1}{L}\sum_{k}e^{i{\bf{k}} \cdot {\bf{R}}_j}G_{k\sigma }^{12}(\omega)d\omega
\label{eq7}.
\end{equation}
 \begin{figure}
     \centering
     \includegraphics[angle=-90,width=10.5cm]{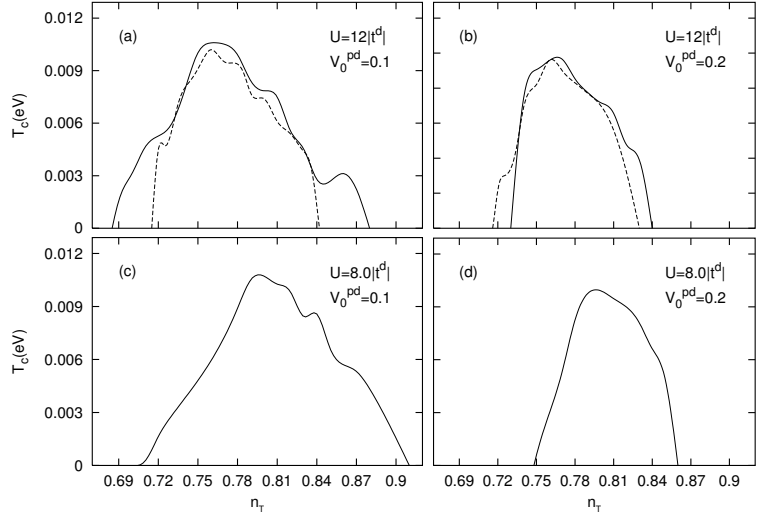}
     \caption{$T_{c}$ as a function of the total occupation number $n_T$
              (where $n_T=n_{\sigma}^d+n_{-\sigma}^d$). In (a) and (b), the dotted
              lines show the previous results from Ref. \cite{ref3} for $U=12|t^d|$.
              The solid lines show the behavior of $T_{c}$ in the present approach.
              The figures (c) and (d) show the present results for $U=8|t^d|$.
              The units of the hybridization $V_0^{pd}$ are
              electron-volts (eV). ($t^d=-0.5eV$.)}
     \label{fig1}
 \end{figure}

The band shift given by equation (\ref{eq5}) can be split into a
$\bf k$-dependent and a $\bf k$-independent term. While the role
of the $\bf k$-independent term is only to shift the poles of the
Green's function, the $\bf k$-dependent part causes a flattening
at the top of the lower band from around ${\bf k}=(\pi,\pi)$ until
${\bf k}=(\pi,0)$ as can be verified in Refs. \cite{ref6,ref8}.
This flattening occurs because $W_{k\sigma}^d$ decreases when
$\varepsilon _{k}^{d}$ increases, as discussed in Ref.
\cite{ref6}. This numerical calculations agree with the Monte
Carlo results obtained by Bulut {\it et al} \cite{ref6,ref8}. The
flattening in the quasi-particle sub-bands (called Hubbard bands)
leads to a band narrowing of these sub-bands, increasing the gap
originated by the Coulomb interaction $U$. Our numerical results
show that the {\it narrowing} of the sub-bands decreases with
increasing $U$. This fact is very important, because this leads to
a change at the position of the chemical potential $\mu$ which is
relevant to obtain the behavior of the high-temperature
superconductivity \cite{ref8}. In the figures
\ref{fig1}(a)-\ref{fig1}(b) are shown phase diagrams with the
dotted lines corresponding to the results obtained in Ref.
\cite{ref3}, where the effects of the temperature $T$ and the
Coulomb interaction are not included in $W_{k\sigma}$. In the
figures \ref{fig1}(c)-\ref{fig1}(d) the value of the $U$ is
decreased. The solid lines show the results for $T_c$, with the
effects of $T$ and $U$ included in the calculation of the
$W_{k\sigma}^d$ band shift. The results show a small increase of
$T_c$ when we consider in $W_{k\sigma}^d$ the effects of $T$ and
$U$. We believe that the increasing of $T_c $ occurs as result of
the change in the value of the chemical potential, as discussed
above. The numerical results show also that in the interval of
temperatures showed in figure \ref{fig1} the effect of the
temperature in $W_{k\sigma}^d$ is quite small.

%
%
%
%

\end{document}